\documentstyle[12pt,epsfig]{article}

\newcommand{\lsim}{\raisebox{-0.13cm}{~\shortstack{$<$ \\[-0.07cm] $\sim$}}~}
\newcommand{\gsim}{\raisebox{-0.13cm}{~\shortstack{$>$ \\[-0.07cm] $\sim$}}~}

\newcommand{\ra}{\rightarrow}

\newcommand{\s}{\smallskip}
\newcommand{\nn}{\noindent}

\newcommand{\beq}{\begin{eqnarray}}
\newcommand{\eeq}{\end{eqnarray}}

\catcode`@=11
\def\citer{\@ifnextchar
[{\@tempswatrue\@citexr}{\@tempswafalse\@citexr[]}}
\def\@citexr[#1]#2{\if@filesw\immediate\write\@auxout{\string\citation{#2}}\fi
  \def\@citea{}\@cite{\@for\@citeb:=#2\do
    {\@citea\def\@citea{--\penalty\@m}\@ifundefined
       {b@\@citeb}{{\bf ?}\@warning
       {Citation `\@citeb' on page \thepage \space undefined}}%
\hbox{\csname b@\@citeb\endcsname}}}{#1}}
\catcode`@=12

\begin{document}

\vspace*{.1cm} 
\baselineskip=15pt

\begin{flushright}
CERN--TH/2003--214\\
PM/03--15\\
October 2003\\
\end{flushright}

\vspace*{0.9cm}

\begin{center}

{\large\sc {\bf PDF uncertainties in Higgs production at hadron colliders}}

\vspace*{7mm}

{\sc Abdelhak DJOUADI}$^{1,2}$ and {\sc Samir FERRAG}$^{3}$ 
\vspace*{5mm} 

$^1$ Theory  Division, CERN, CH--1211 Geneva 23, Switzerland.
\vspace*{2mm}

$^2$ Laboratoire de Physique Math\'ematique et Th\'eorique, UMR5825--CNRS,\\
Universit\'e de Montpellier II, F--34095 Montpellier Cedex 5, France. 
\vspace*{2mm} 

$^3$ LPNHE--Paris, Universit\'e Paris 6 et 7 and IN2P3 CNRS, France.
\end{center} 

\vspace*{1cm} 

\begin{abstract}

\nn Using the new schemes provided by the CTEQ and MRST collaborations and by
Alekhin, we analyse the uncertainties due to the parton distribution functions
(PDFs) on the  next-to-leading-order cross sections of  the four main
production processes of the Standard Model Higgs boson at the LHC and the
Tevatron. In the Higgs mass range where the  production rates are large
enough, the spread in the uncertainties when the three sets of PDFs are
compared is of about 15\% in all processes and at both colliders. However,
within one given set of PDFs, the deviations from the values obtained with the
reference sets are much smaller, being of ${\cal O}(5$\%), except in the
gluon--gluon fusion mechanism at relatively large Higgs boson masses, where
they can reach the level of 10\% (15\%) at the LHC (Tevatron).  

\end{abstract}

\newpage

The discovery of the Higgs boson is the ultimate test of the Standard Model of
the electroweak interactions, and the search for this particle is the major
goal of the next round of high-energy experiments. If the Higgs boson is
relatively light, $M_H \lsim 200$ GeV, as is suggested by the electroweak
precision measurements \cite{EW}, it can be produced at the Tevatron Run II if
enough integrated luminosity is collected \cite{Tevatron,Houches}.  At the LHC,
the Higgs boson can be produced over its entire mass range, $M_H \lsim {\cal
O}(1~{\rm TeV})$, in many and sometimes redundant channels \cite{Houches,LHC}. 
Once the Higgs boson is found, the next step would be to perform accurate
measurements to explore all its fundamental properties. To achieve this goal in
great detail,  all possible Higgs cross sections and decay branching  ratios
should be measured in the most accurate manner. At the same time, we need
very precise predictions and a good estimate of the various theoretical
uncertainties that still affect these production cross sections and decay
branching ratios, once higher effects are included. \s

Parton distribution functions (PDFs), which describe the momentum distribution
of a parton in the proton,  play a central role at hadron colliders.  A precise
knowledge of the PDFs over a wide range of the proton momentum fraction $x$
carried by the parton and the squared centre-of-mass  energy $Q^2$ at which the
process takes place, is mandatory to precisely predict the production cross
sections of the various signals and background hard processes. However, they
are plagued by uncertainties, which arise either from the starting
distributions obtained from a global fit  to the available data from
deep-inelastic scattering, Drell--Yan and hadronic data, or from the DGLAP
evolution  \cite{DGLAP} to the higher $Q^2$ relevant to the Tevatron and LHC
scattering processes.   Together with the effects of unknown perturbative
higher order corrections, these uncertainties dominate the theoretical error on
the predictions of the production cross sections. \s

PDFs with intrinsic uncertainties became available in 2002. Before that date,
to quantitatively estimate the uncertainties due to the structure functions, it
was common practice to calculate the production cross sections using the
``nominal fits" or reference set of the PDFs provided by different
parametrizations and to consider the dispersion between the various predictions
as being the ``uncertainty" due to the PDFs. However, the comparison between
different parametrizations cannot be regarded as an unambiguous way to estimate
the uncertainties since the theoretical and experimental errors spread into
quantitatively different intrinsic uncertainties following  their treatment in
the given parametrization. The CTEQ and MRST collaborations and Alekhin
recently introduced new schemes, which provide the possibility of estimating
the  intrinsic uncertainties and the spread  uncertainties on the prediction 
of physical observables at hadron colliders\footnote{Other sets of PDFs with
errors are available in the literature \cite{otherPDF}, but they will not be
discussed here.}. \s

In this note, the spread uncertainties on the Higgs boson production cross
sections at the LHC and at the Tevatron, using the CTEQ6 \cite{CTEQ6}, 
MRST2001 \cite{MRST2001E} and ALEKHIN2002 \cite{ALEKHIN} sets of PDFs, 
are investigated and compared. \s 

The scheme introduced by both the CTEQ and MRST collaborations is based on the
Hessian matrix method.  The latter enables a characterization of a parton
parametrization in the neighbourhood of the global $\chi^2$ minimum fit and 
gives an access to the uncertainty estimation through a set of PDFs that
describes this neighbourhood. Fixed target Drell--Yan data as well as $W$
asymmetry and jet data from the  Tevatron are used in the fit procedure. 

The corresponding PDFs are constructed as follows: 

\begin{itemize}
\vspace*{-2mm}

\item[--] a global fit of the data is performed using the free parameters
$N_{\rm PDF}=20$ for CTEQ and $N_{\rm PDF}=15$ for MRST; this provides the
nominal PDF (reference set) denoted by $S_0$ and corresponding to CTEQ6M and
MRST2001E, respectively;  \vspace*{-3mm}

\item[--] the global $\chi^2$ of the fit  is increased by $\Delta \chi^2=100$
for CTEQ and $\Delta \chi^2=50$ for MRST, to obtain the error matrix [note that
the choice of an allowed tolerance is only intuitive for a global analysis
involving a number of different experiments and processes];  \vspace*{-3mm}

\item[--] the error matrix is diagonalized to obtain $N_{\rm PDF}$ eigenvectors
corresponding to $N_{\rm PDF}$ independent directions in the parameter space; 
\vspace*{-3mm}

\item[--] for each eigenvector, up and down excursions are performed in the 
tolerance gap, leading to $2N_{\rm PDF}$ sets of new  parameters, corresponding
to 40 new sets of PDFs for CTEQ and 30 sets for MRST. They are denoted by $S_i$,
with $i=1, 2N_{\rm PDF}$.   \vspace*{-2mm} 
\end{itemize} 

To built the Alekhin PDFs \cite{ALEKHIN}, only light-target  deep-inelastic
scattering data [i.e. not the Tevatron data] are used. This PDF set involves 14
parameters, which are fitted simultaneously with $\alpha_s$ and the structure
functions.  To take into account the experimental errors and their
correlations,  the fit is performed by minimizing a $\chi^2$ functional based
on a covariance matrix.  Including the uncertainties on the $\alpha_s$ fit, one
then obtains $2N_{\rm PDF}=30$  sets of PDFs for the uncertainty estimation. \s

The three sets of PDFs discussed above are used to calculate the uncertainty on
a cross section $\sigma$ in the following way \cite{Samir}: one first evaluates
the cross section with the nominal PDF $S_0$ to obtain  the central value
$\sigma_0$. One then calculates the cross section with  the $S_i$ PDFs, giving 
$2N_{\rm PDF}$ values $\sigma_i$, and defines, for each $\sigma_i$ value, the 
deviations  $\Delta \sigma_i^\pm =\mid \sigma_i -\sigma_0\mid$ when $\sigma_i \
^{>}_{<}  \sigma_0$. The uncertainties are summed quadratically to calculate
{\bf $\Delta  \sigma^\pm  = \sqrt{ \sum_i \sigma_i^{\pm 2} }$}.  The cross
section, including the error, is then given by $\sigma_0|^{+\Delta \sigma^+}_{-
\Delta \sigma^-}$.   \s

This procedure is applied to estimate the cross sections for the production of
the Standard Model Higgs boson in the following four main mechanisms \cite{LO}:
\beq
{\rm associate~production~with}~W/Z: & & q\bar{q} \ra VH \\
{\rm massive~vector~boson~fusion}: & & qq \ra   Hqq \\
{\rm the~gluon~gluon~fusion~mechanism}: & & gg  \ra H \hspace*{2cm} \\
{\rm associate~production~with~top~quarks}: & & gg,q\bar{q}\ra t \bar{t} H
\eeq

We  will consider the NLO cross sections for the production at both the LHC and
the Tevatron, and use the Fortran codes {\tt V2HV, VV2H, HIGLU} and {\tt  HQQ}
of Ref.~\cite{Michael} for the evaluation of the production cross sections of 
processes (1) to (4), respectively. A few remarks are to be made in this
context: 

\begin{itemize}
\vspace*{-2mm}

\item The NLO QCD corrections to the Higgs-strahlung processes
\cite{NLOHV,M+A}  are practically the same for $WH$ and $ZH$ final states; we
thus simply concentrate on the process $q\bar{q} \to WH$, which has a larger
cross section at the LHC and at the Tevatron.   
\vspace*{-2mm}

\item The vector boson fusion process, $pp \to Hqq$, for which the NLO
corrections have been  calculated in \cite{NLOVV,M+A,Dieter}, is only relevant
at the LHC  and will not be discussed in the case of the Tevatron, where the
cross sections are too small to be relevant.   \vspace*{-2mm}

\item For the gluon fusion process, $gg \to H$, we include the full  dependence
on the top and bottom quark masses of the NLO cross section \cite{NLOggfull} 
and not only the result in the infinite top  quark mass limit \cite{NLOgg}.  
\vspace*{-2mm}

\item For the $pp \to Ht\bar{t}$  production process, the NLO corrections have
been calculated only recently \cite{NLOHtt} and the programs [which are very
slow because of the complicated final state] for calculating the cross sections
are not yet publicly available.  However, we choose a scale for which the LO
and NLO cross sections are approximately equal and use the program {\tt HQQ}
for the LO cross section that we fold with the NLO PDFs. 
\vspace*{-2mm}

\end{itemize}

Finally, we note that the NNLO corrections are also known in the case of the
Higgs-strahlung $q\bar{q} \to HV$  \cite{NNLOHV} and fusion $gg \to H$ [in the
infinite top quark mass limit] \cite{NNLOgg,Catani} processes. We do not
consider these higher order corrections since the CTEQ and MRST PDFs with
errors are not available at this order\footnote{In fact, even the nominal PDFs
are not  known completely at NNLO since the full Altarelli--Parisi    splitting
functions are not yet available at this order of perturbation theory.  However,
the MRST collaboration and Alekhin have approximate solutions; only the Alekhin
set includes uncertainty estimates, though.}. The errors for the $gg\to H$
process at NNLO, including soft-gluon resummation, have been discussed in
Ref.~\cite{Catani} using an approximate NNLO PDF set provided by Alekhin
\cite{ALEKHIN}. \s

The expected NLO Higgs boson production cross sections at the LHC and the
Tevatron, as a function of the Higgs mass, are shown in Fig.~1, using the
CTEQ6M reference set for the PDFs.  As can be seen, at the LHC, the  cross
sections for $gg \to H$ and $qq \to qqH$ are above the 0.1 pb level for Higgs
masses up to 1 TeV, while for the $q\bar{q} \to HW$ and $q\bar{q}/gg \to
t\bar{t}H$ processes they become of the order of 0.1 pb for Higgs masses around
200 GeV.  At the Tevatron, only the processes $gg \to H$ and $q\bar{q} \to HW$
have  sizeable cross sections [$\gsim 0.01$--0.1 pb] for Higgs boson masses
below 200 GeV. We will therefore simply concentrate on these particular
processes in the Higgs boson mass range where they are relevant. \s

\begin{figure}[hbtp]
\begin{center}
\vspace*{-3cm}
\hspace*{-2cm}
\psfig{figure=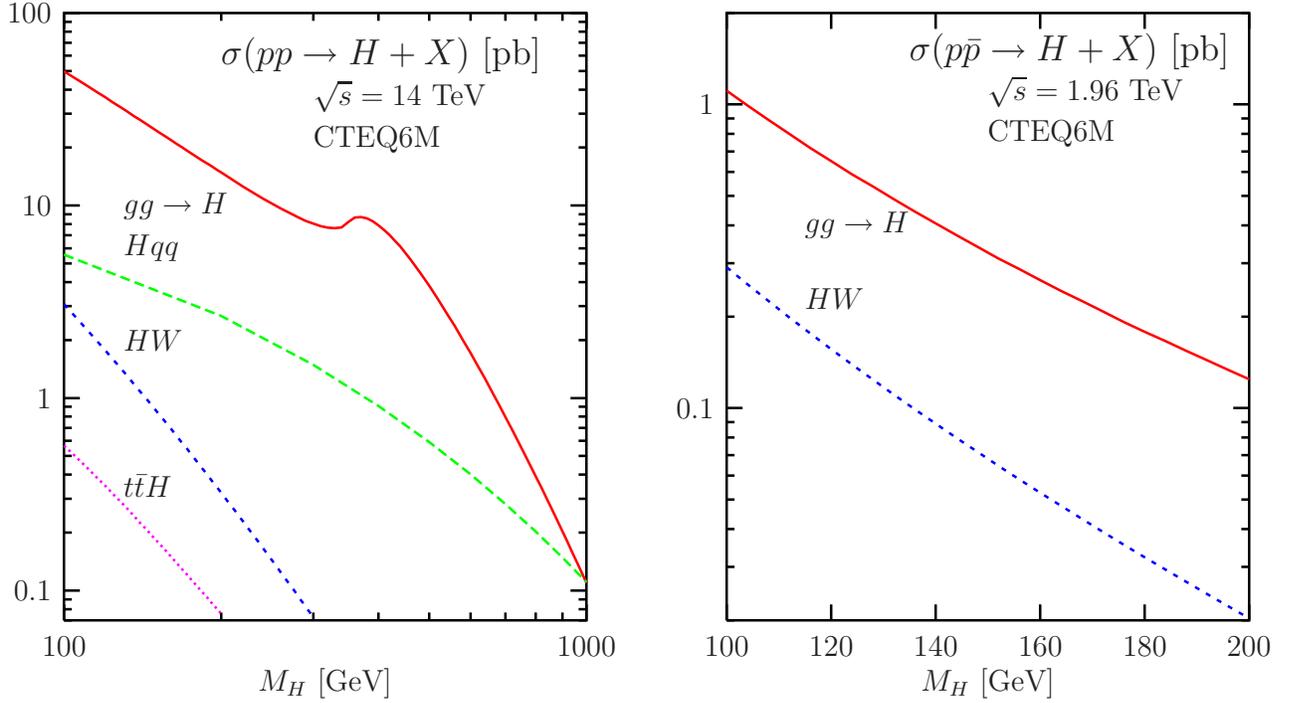,width=20cm}
\vspace*{-17.6cm}
\end{center}
\caption[]{\it The NLO cross sections for Higgs production at the LHC (left)
and the Tevatron (right) as a function  of the Higgs mass. The reference CTEQ6M
set is used.}
\vspace*{-3mm}
\end{figure}

Before analysing the uncertainties on the production cross sections, let us
first discuss and  compare the various PDFs. In Fig.~2, the MRST and Alekhin
densities for the gluon and for the up and down quarks and antiquarks,
normalized to the CTEQ6 ones, are displayed for a wide range of $x$ values and
for a fixed c.m. energy $Q^2=(100\ {\rm GeV})^2$. One notices the following
main features [the same gross features are observed for $Q^2=(500 \ {\rm
GeV})^2$]:

\begin{itemize}
\vspace*{-2mm}

\item[--] The MRST gluon PDF is smaller than the CTEQ one, except for $x$
values  around $x\sim 0.1$; in contrast, the Alekhin gluon PDF is larger than
the  CTEQ one for all $x$ values, except for values of $x$ around $x \sim 0.01$
and for very high $x$.   \vspace*{-2mm}

\item[--] The MRST (anti)quark PDFs are practically equal in magnitude and are
smaller than the CTEQ ones for low $x$ values, while they are in general
slightly larger for higher $x$, except for values near unity; in the Alekhin
case, all (anti)quark PDFs are larger than the CTEQ ones, except for the
$\bar{u}$ density above $x \sim 0.05$. For values, $x \gsim 10^{-4}$, the
differences between the Alekhin and the CTEQ6 PDFs are more pronounced than the
differences between the MRST and the CTEQ ones.  \vspace*{-2mm} \end{itemize}

\begin{figure}[hbtp]
\begin{center}
\vspace*{-3cm}
\hspace*{-2cm}
\psfig{figure=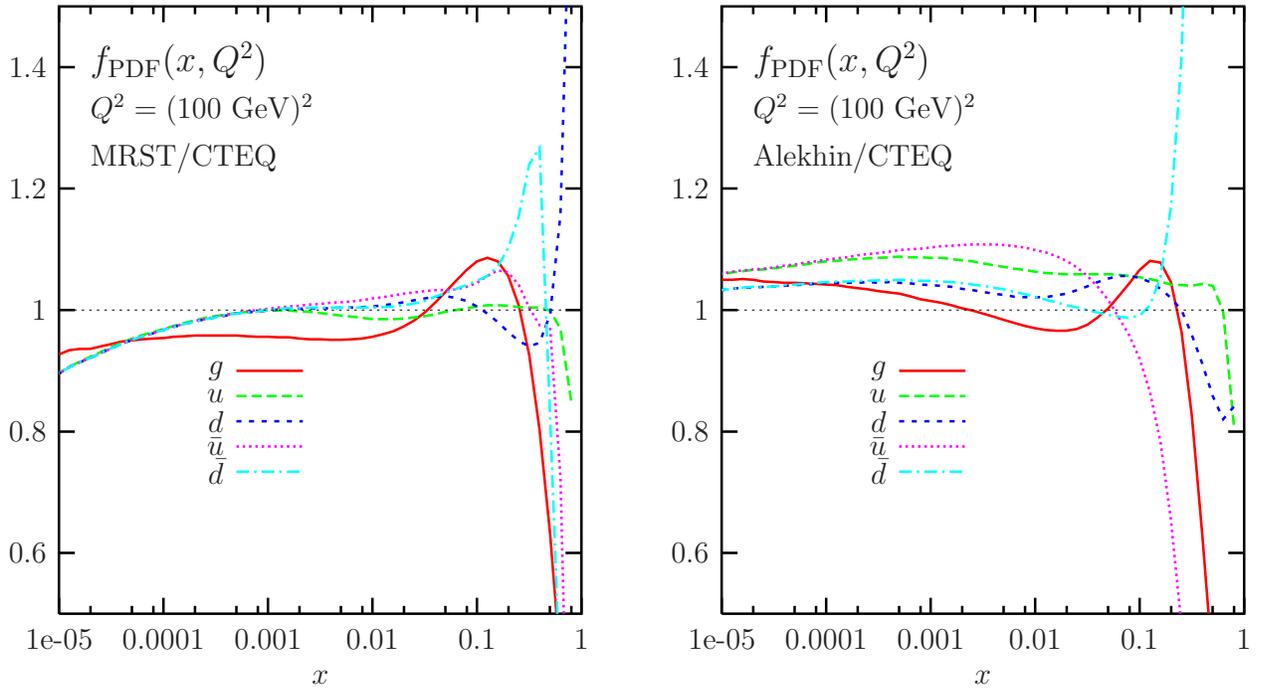,width=20cm}
\vspace*{-17.6cm}
\end{center}
\caption[]{\it MRST and Alekhin densities for the gluon, up quark/down
quark and antiquarks, normalized to the CTEQ6 ones, as a function of $x$ and
for $Q^2=(100\ {\rm GeV})^2$.}
\vspace*{-3mm}
\end{figure}

Let us now comment on the intrinsic uncertainties of the PDF sets of the CTEQ 
and MRST collaborations, which follow the same approach. As discussed in  
Refs.~\cite{CTEQ6} and \cite{MRST2001E},  three different behaviours of the
uncertainty bands can be distinguished,  according to three different ranges of
the variable $x$: decreasing uncertainties at low $x$, constant or slightly
oscillating ones at intermediate $x$, and increasing  ones at high $x$. The
magnitude of these uncertainties depends on the considered parton  and on the
c.m. energy $Q^2$. In the case of quarks, the three behaviours are observed:
the low-$x$ behaviour extends up to $x \sim$ few $10^{-3}$, and the high-$x$
one starts in the neighbourhood of $x=0.7$. At high $Q^2$, the uncertainties at
high and low-$x$ values exceed a few tens of a per cent and in the intermediate
regime, they are less than a few  per cent. In the gluon case and at high
$Q^2$, the low-$x$ and the intermediate-$x$ bands are not as well separated as
in the case of quarks; the uncertainty band reaches also the few per cent
level. The high-$x$  regime starts in the neighbourhood of $x \sim 0.3$, i.e
earlier than in the case of quarks.  \s

The behaviour of the Higgs production cross sections and their uncertainties
depends on the considered partons and their $x$ regime discussed above.  In
Figs.~3 and 4, we present the cross sections in, respectively,  the case of the
$q\bar{q} \to HW$ and $gg \to H$  processes at both the LHC and the Tevatron,
while in Fig.~5,  we show the cross sections in the case of the $qq \to qqH$
and $pp \to ttH$ processes at the  LHC only. The  central values and the 
uncertainty band limits of the NLO cross sections are shown for the CTEQ, MRST
and Alekhin parameterizations. In the insets to these figures, we show
the spread uncertainties in the predictions for the NLO cross sections, when
they are normalized to the prediction of the reference CTEQ6M set. Note that
the three sets of PDFs do not use the same value for  $\alpha_s$: at NLO, the
reference sets CTEQ6M, MRST2001C and A02 use, respectively, the values
$\alpha_s^{\rm NLO}(M_Z)= 0.118$, $0.119$  and 0.117. \s

By observing Figs.~3--5, we see that the uncertainties for the Higgs  cross
sections obtained using the CTEQ6 set are two times larger than those using the
MRST2001 sets. This is mainly due to two reasons: first, as noted previously,
the CTEQ collaboration increased the global $\chi^2$ by $\Delta\chi^2=100$ to
obtain the error matrix, while the  MRST collaboration used only
$\Delta\chi^2=50$; second, 2$\times$20 parameter uncertainties are summed
quadratically in CTEQ6, while only 2$\times$15 are used in the MRST case. The
uncertainties from  the Alekhin PDFs are larger than the MRST ones and smaller
than the CTEQ ones. In the subsequent discussion,  the magnitude of the
uncertainty band is expressed  in terms of the CTEQ6 set.  \s

$\bullet$ $q\bar{q} \ra VH$: at the LHC, the uncertainty band is almost
constant  and is of the order of 4\% [for CTEQ] over a Higgs masse range
between 100 and 200 GeV.  At the Tevatron, the uncertainty band increases with
the Higgs mass and exceeds 6\% at $M_{H}\sim 200$ GeV.  To produce a vector
plus a Higgs boson in this mass range,  the incoming quarks originate from the
intermediate-$x$ regime at the LHC, at Tevatron energies, however, some of the
participating quarks originate from the high-$x$ regime. This explains the
increasing  behaviour of the uncertainty bands observed in the Tevatron case. 
The  different magnitude of the cross sections, $\sim 12$\% ($\sim 8$\%) larger
in the Alekhin case than for CTEQ  at the LHC (Tevatron), is due to the larger
quark and antiquark densities; see Fig.~2. For this particular PDF set, the
difference in the shifts of the central values in the LHC and Tevatron cases, 
is due to the different initial states at the two machines: in $pp$ collisions,
the antiquark comes from the sea, while in  $p\bar{p}$ collisions, it is a
valence+sea antiquark  and the sea quark shift compared to the CTEQ case is
more important than the valence+sea one; see Fig.~2.  \s

\begin{figure}[hbtp]
\begin{center}
\vspace*{-3cm}
\hspace*{-2cm}
\psfig{figure=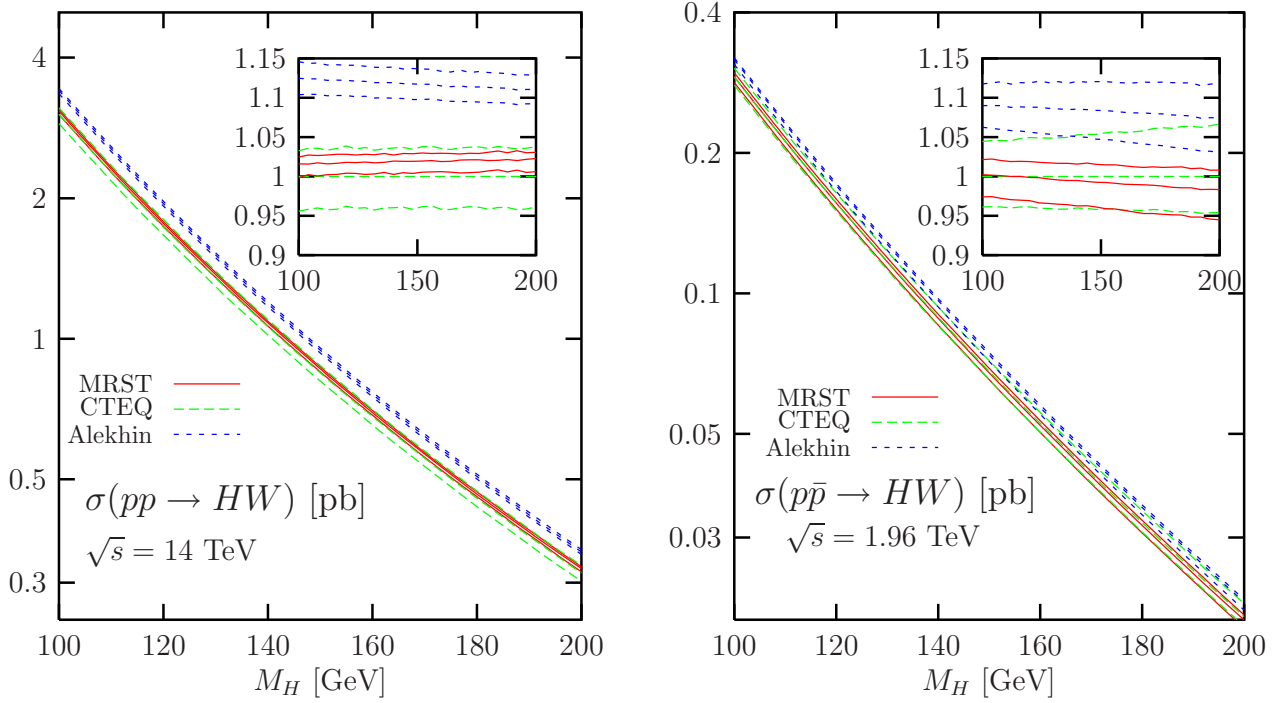,width=20cm}
\vspace*{-17.5cm}
\end{center}
\caption[]{\it The CTEQ, MRST and Alekhin PDF uncertainty bands for the NLO
cross sections for the production of the Higgs boson at the LHC (left) and at
the Tevatron (right) in the $q\bar{q} \to HW$  process.  The insets 
show the spread in the predictions, when the NLO cross sections are
normalized to the prediction of the reference CTEQ6M set.}
\vspace*{-3mm} 
\end{figure}

\begin{figure}[hbtp]
\begin{center}
\vspace*{-3cm}
\hspace*{-2cm}
\psfig{figure=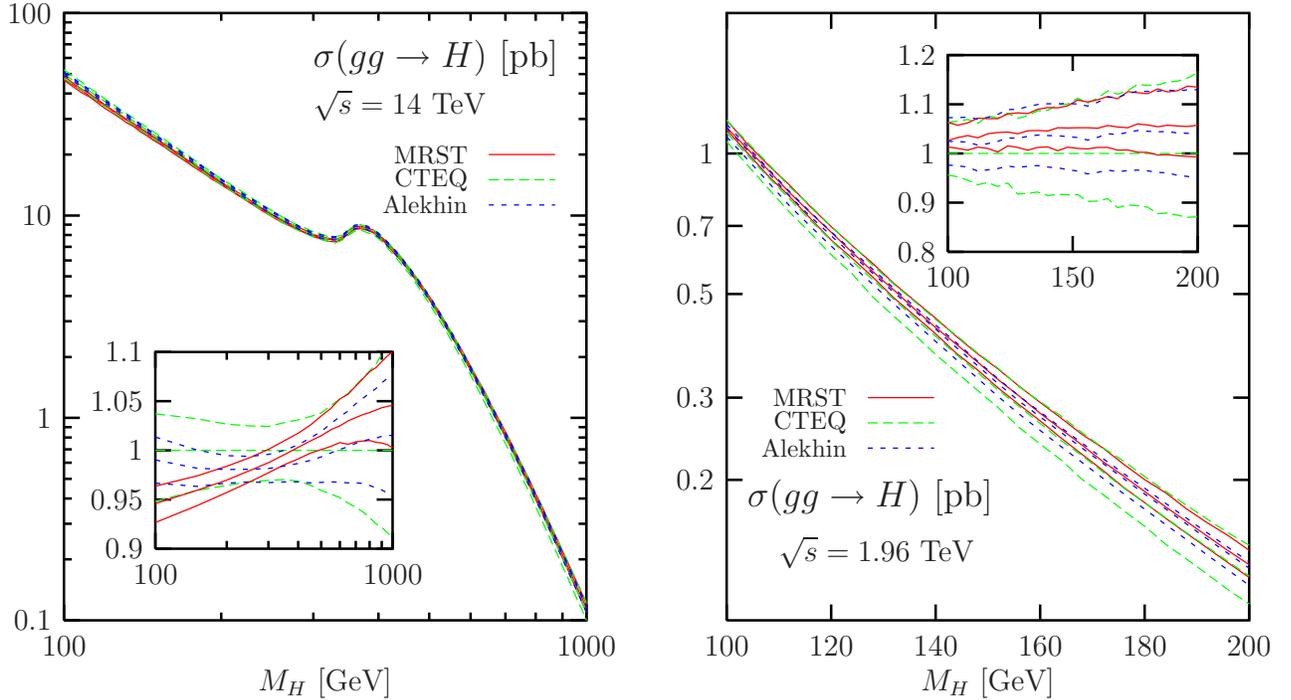,width=20cm}
\vspace*{-17.2cm}
\caption[]{\it Same as Fig.~3, but for the  $gg \to H$ production process.}
\end{center}
\vspace*{-7mm}
\end{figure}

\begin{figure}[hbtp]
\begin{center}
\vspace*{-3cm}
\hspace*{-2cm}
\psfig{figure=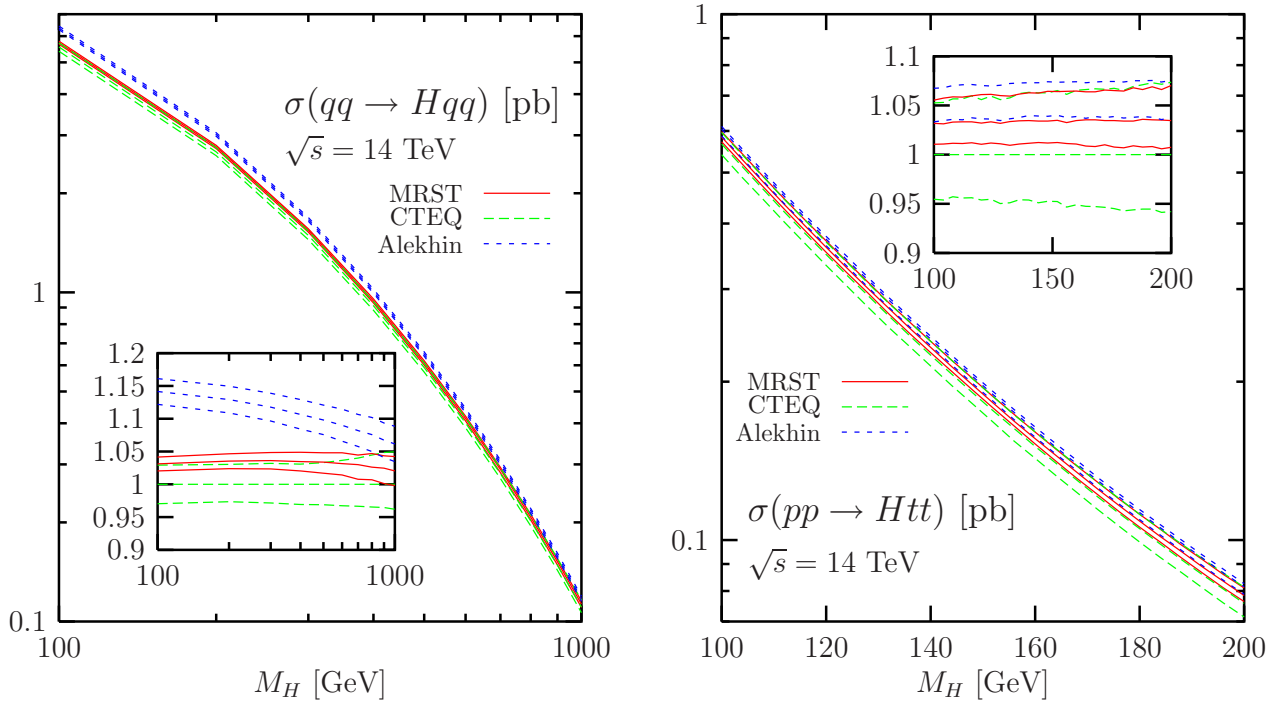,width=20cm}
\end{center}
\vspace*{-17.2cm}
\caption[]{\it Same as Fig.~3, but for the  $qq \to Hqq$  and $pp \to
ttH$  processes at the LHC.}
\vspace*{-5mm}
\end{figure}

$\bullet$ $gg  \ra H$: at the LHC, the uncertainty band for the CTEQ set of
PDFs  decreases from the level of about 5\% at $M_{H} \sim 100$ GeV, down to
the 3\% level at $M _H \sim$ 300 GeV.  This is because Higgs bosons with
relatively small masses  are mainly  produced by  asymmetric  low-$x$--high-$x$
gluons  with a low effective c.m. energy; to produce heavier Higgs bosons, a
symmetric process in which the participation of intermediate-$x$ gluons with
high density, is needed, resulting in a smaller  uncertainty band. At higher
masses, $M_H \gsim 300$ GeV, the participation of  high-$x$ gluons becomes more
important, and the uncertainty band increases, to reach the 10\% level at 
Higgs masses of about 1 TeV. At the Tevatron, because of  the smaller c.m.
energy, the high-$x$ gluon regime is already reached for low Higgs masses and
the uncertainties increase from 5\% to 15\% for $M_H$ varying between 100 GeV
and 200 GeV. As discussed above and shown in Fig.~2, the MRST  gluon PDF is
smaller than the CTEQ one for low $x$ and larger for relatively high $x$ ($\sim
0.1$): this explains the increasing cross section obtained with MRST compared
to the one obtained with CTEQ, for increasing Higgs boson mass  at the LHC.  At
the  Tevatron the gluons are already in the high-$x$ regime. \s

$\bullet$ $gg/q\bar{q}\ra t\bar{t}H$: at the LHC, the associated production of
the Higgs boson  with a top quark pair is dominantly generated by the 
gluon--gluon fusion mechanism. Compared with the  process $gg  \ra H$ discussed
previously  and for a fixed Higgs boson mass, a larger $Q^2$ is needed for this
final state;  the initial gluons should therefore have higher $x$ values. In
addition, the quarks that are involved in the subprocess $q\bar{q}\ra
t\bar{t}H$, which is also contributing,  are still in  the intermediate regime
because of the higher value $[x \sim 0.7$] at which the quark high-$x$ regime
starts.  This explains why the uncertainty band increases smoothly from 5\% to
7\% when the $M_H$ value increases from 100 to 200 GeV.\s

$\bullet$ $qq \ra   Hqq$: in the entire Higgs boson mass range from 100 GeV  to
1 TeV, the incoming quarks involved in this process originate from the
intermediate-$x$ regime and the uncertainty band is almost constant, ranging
between 3\% and 4\%. [This behaviour agrees with the one discussed in
Ref.~\cite{Dieter}, where a uniform 3.5\% uncertainty using the CTEQ PDF has
been found.] When using the Alekhin  set of PDFs, the behaviour is different, 
because the quark PDF behaviour is different, as discussed in the case of the
$q\bar{q}  \to HV$  production channel. The decrease in the central value with
higher Higgs boson mass [which is absent in the $q\bar{q} \to HV$ case, since
we stop the $M_H$ variation at 200 GeV] is due to the fact that we reach here
the high-$x$ regime, where the Alekhin $\bar{u}$ PDF drops steeply. \s

Finally, it should be noted that, besides the uncertainties on the PDFs 
discussed here, which can be viewed as ``experimental uncertainties" since they
concern the systematic and statistical uncertainties of the data included in
the global fits for a given set of PDFs,  there are several  other sources of
uncertainties on the PDFs, which are associated with the global parton
analysis, which can be viewed as ``theoretical errors".  Among these are the
uncertainties due to the input assumptions, the selection of the fitted data,
the truncation of the  DGLAP perturbative series, and theoretical effects such
as higher twist effects, etc.  The impact of these errors, for instance in the
case of the MRST PDF set, has been discussed recently \cite{THerrors}.  The
discussion of these errors is beyond the scope of the present note. However, in
our analysis, we have used three different sets of PDFs in which many of the
previous items are  treated differently.  One could, therefore, consider the
spread in the predictions given by the three (reference) sets of PDFs as a
rough measure of these theoretical uncertainties. \s

In summary,  we have considered three sets of PDFs with uncertainties provided
by the CTEQ and MRST collaborations and by Alekhin. We evaluated their impact
on the total cross sections at next-to-leading-order for the production of the
Standard Model Higgs boson at the LHC and at the Tevatron. Within a given set
of PDFs, the deviations of the cross sections from the values obtained with the
reference PDF sets are rather small, ${\cal O}(5$\%), in  the case of the
Higgs-strahlung, vector boson fusion and associated  $t\bar{t}H$ production
processes, but they can reach the level of 10\% (15\%) at the LHC (Tevatron) in
the case of the gluon--gluon fusion process for large enough Higgs boson
masses, $M_H \sim 1$ TeV ($\sim 180$ GeV). However, the relative differences
between the cross sections evaluated with different sets of PDFs can be much
larger. Normalizing to the values  obtained with the CTEQ6M set, for instance, 
the cross sections can be different by up to 15\% for the four production
mechanisms. \bigskip

\nn {\bf Acknowledgments:} We thank S. Alekhin, J. Huston and M. Spira for 
discussions. This work has been initiated during the Les Houches Workshop and
we thank the organisers.

\end{document}